\documentclass{edp-jp4}
\usepackage{graphicx}

\def\etal{{\it et al.} }

\def\apj{{\it ApJ}}

\def\aa{{\it A\&A}}

\def\mjysr{MJy/sr }

\def\inufit{I_{\nu fit}}

\def\bnu{{B_{\nu}}}
\def\msol{{M$_{\odot}$}}
\def\mic{{\mu}m}
\def\cm2{$cm^{-2}$}

\begin{document}

\title{Dust emission in massive star-forming regions with PRONAOS: the Orion and M17 molecular clouds}

\author{Xavier Dupac}\address{Centre d'\'Etude Spatiale des Rayonnements - CESR, 9 av. du Colonel Roche, BP 4346, F-31028 Toulouse Cedex 4, France 
}
\author{the PRONAOS collaboration}

\maketitle

\begin{abstract} The balloon-borne submillimeter instrument PRONAOS has
  observed one square degree areas towards the Orion and M17 molecular
  clouds. The 2' - 3.5' resolution
maps obtained in four wide wavelength bands between 200 $\mic$ and 600 $\mic$, exhibit the dust distribution in these regions. We analyze the temperature and
spectral index of the dust, and we show the anticorrelation between these two
  parameters. We derive estimations of the ISM column densities and masses in these regions. 
\end{abstract}

\section{Introduction}
The high-mass star-forming regions of our Galaxy are known to be the site of
complex physical and chemical processes, particularly involving dust grains.
The submillimeter wavelength domain is particularly interesting to characterize dust
properties.
Dust emission in this spectral range is mainly due to big grains at thermal
equilibrium (see, e.g., D\'esert \etal 1990), whose emission is usually modelled
by the modified blackbody law, introducing the temperature and the spectral
index of the dust.
The temperature of a molecular cloud is a key parameter which controls
(with others) the structure and evolution of the clumps, and therefore, star formation.
Thus spectral imaging of molecular clouds can provide a large amount of
knowledge about their structure and evolution, especially if the dust emission
parameters are able to be properly derived on top of submillimeter intensities.
Mapping of star-forming molecular clouds, as well as other dusty regions, has
been performed by the PRONAOS balloon-borne experiment (PROgramme NAtional d'Observations
Submillim\'etriques, see Buisson \& Duran 1990 or Ristorcelli \etal 1998).
We present here some results obtained on Orion and M17.
The full analysis on Orion can be found in Dupac \etal 2001.

\section{Observations and data processing}
PRONAOS (PROgramme NAtional d'Observations Submillim\'etriques) is a French
balloon-borne submillimeter experiment, with a 2 m diameter telescope
(Buisson \& Duran 1990).
The focal plane instrument SPM (Syst\`eme Photom\'etrique Multibande, see
Lamarre \etal 1994) is
composed of a wobbling mirror, providing a beam switching on the sky with an
amplitude of about 6' at 19.5 Hz, and four bolometers cooled at 0.3 K. They
measure the submillimeter flux in the spectral ranges 180-240 $\mic$, 240-340
$\mic$, 340-540 $\mic$ and 540-1200 $\mic$, with sensitivity to low
brightness gradients of about 1 \mjysr in band 4.
The angular resolutions are 2' in bands 1 and 2, 2.5' in band 3 and 3.5' in
band 4.
The data which we present here were obtained during the second flight of PRONAOS in
september 1996, at Fort-Sumner, New Mexico.
The observing procedure is an altazimut scanning of the beam on the
sky.
The first processing applied to the data is correction from map
distortion, taking into account the pointing of the telescope, including fine
pointing errors due to the swinging of the gondola.
Then we make the maps with the method described in Dupac \etal (2001),
which is based on direct linear
inversion on the whole map, using a Wiener matrix.
This map-making process takes into account the beam sizes and profiles, the
beam switching and the signal and noise properties to construct an optimal
map.
For the M17 data, we have used an improved method, in which we
consider some noise not independent from the sky signal.

\section{Results and analysis}
We present in Fig. 1 the images of Orion M42 and M17 obtained in the two extreme photometric bands of
PRONAOS-SPM.
The noise level is about 4 \mjysr rms in band 1 and 0.8 \mjysr in band 4.
However, due to the calibration uncertainty, the flux accuracy is not better
than 5 \% (1 $\sigma$) relative between bands (10-20 \% absolute).

\begin{figure}[]
\caption[]{PRONAOS maps of Orion (top) and M17 (bottom) in band 1 (200 $\mic$,
  top of each) and band 4 (580 $\mic$,
  bottom of each).
The color scale is logarithmic, and displays the positive reconstructed flux
until -1.5 in log (Orion) or -1 (M17), then the deeper blue and purple colors display the negative
noise features.
The black lines in the color bars show the contour levels.
The black box drawn in the 580 $\mic$ Orion map represents the area mapped by Ristorcelli \etal (1998).
}
\end{figure}

These maps enhance the very important intensity contrasts which exist in this
kind of
giant molecular complexes.

We assume the emission of the grains to obey the modified blackbody law:

$\inufit(\lambda,T,n)=C . \bnu(\lambda,T) . \lambda^{-\beta}$

where $\lambda$ is the wavelength, C a constant, T the temperature of the grains, $\beta$ the spectral
index and $\bnu$ the Planck function.

The three parameters C, T and $\beta$, are adjusted with a least square fit.

From a detailed analysis of the temperature and spectral index spatial
distribution in these
molecular clouds, we deduce the following main results.
The temperature varies roughly from 10 K to 80 K, the coldest dust being in
the outskirts of the star-forming complex, mainly in the form of cold
(early protostellar ?) clumps.
The spectral index varies roughly from 1 to 2.5, the highest values being
observed in the coldest areas.
Indeed, it seems that exists an anticorrelation
between the temperature and the spectral index.
The correlation coefficient that we get in Orion is -0.9, and we obtain -0.8 in M17.
Actually, the fit made on the PRONAOS submillimeter measurement induces an
artificial amount of correlation, which is due to the relative insensitivity
to spectral index variations of low-temperature submillimeter measurements,
and to the relative insensitivity to temperature variations of
high-temperature measurements.
However, this artifact is clearly not enough to explain the anticorrelation
found on the data.
Therefore, this anticorrelation has to be an intrinsic property of the grains,
or at least a property of the observed dust, when integrated on
the ISM column.
From further considerations, we rather support a fundamental explanation for
this anticorrelation effect, involving the physical properties of the dust
grains.
Laboratory experiments (see Agladze \etal 1996 and Mennella \etal 1998) showed this effect on
grains for temperatures
down to 10 K.
Agladze \etal (1996) measured absorption spectra of crystalline and amorphous
grains between 0.7 and 2.9 mm wavelength. They deduced an anticorrelation
between the power-law index $\beta$ and the temperature in the temperature range
10-25 K, and attributed it to two level tunnelling processes.
The measures of Agladze \etal are insufficient to justify
our observation in the submillimeter spectral range, because absorption can be
very different than in the millimeter range.
Mennella \etal (1998) measured the absorption coefficient of cosmic dust
analogue grains, crystalline and amorphous, between 20 $\mic$ and 2 mm
wavelength, in the temperature range 24-295 K. They deduced an
anticorrelation between T and $\beta$, and attributed it to two phonon
difference processes.
On our observations, we observe this effect down to about 15 K, thus
we would need laboratory results on these low temperatures in the
submillimeter range to fully understand the observations.

We also derived estimations of the column densities and masses in Orion and
M17.
For this we use
the dust 100 $\mic$ opacity from D\'esert \etal (1990).
It allows us to
estimate the column density from the spectral intensity. We consider
only the thermal emission of the big grains, which dominate widely in this
spectral range, and we assume that the spectral index does not change in the
PRONAOS spectral range. We take into account the variability of the dust
spectral index of the different regions.
This gives us a simple self-consistent
model that allows us to estimate the column density $N_{H}$, as a function of
the spectral intensity and the spectral index.
We adopt the value of the
opacity $\kappa_{100 \mic}$ = 0.361 $cm^{2}/g$ (per gram of total medium: gas
and dust) at 100 $\mic$, because this value is well constrained from IRAS
data.
This is calculated from the extinction curves given in D\'esert \etal
The opacity $\kappa$ is defined as $\tau\over{N_H.m_H}$, thus the relation
between the gas column density and the fit parameters is:

$N_{H} = {C . \lambda^{-\beta}\over\kappa . m_H}$

where $m_H$ is the proton mass and $\kappa$ the gas and dust opacity.
Then for the D\'esert \etal value of $\kappa_{100 \mic}$ we have:

$N_{H}$ = 1.67 $10^{24}$ . C . (100  $\mic)^{-\beta}$

where $N_{H}$ is the total column density in mass of the gas, in
protons/cm$^{2}$, and C is in unit of $\mic^{\beta}$.
Then we compute the masses of the studied regions, assuming a distance of 470
pc (Orion) and 2200 pc (M17), by integrated the column density on the surface
of the cloud.
For example, we derive a total mass of the M17 complex of 18000 \msol.
Some cold clouds that we observe in Orion and M17, near the warm star-forming area, could be gravitationally
unstable.

As a small conclusion, we can say that these observations could be sustained by sensitive continuum
observations in the millimeter domain, especially for
the faint clouds evidenced, in order to better
constrain the spectral index measurement of the cold dust.
Also, infrared observations could be useful, particularly of the
possible embedded protostars in those faint clouds.

\section{Acknowledgements}
We are indebted to the French space agency Centre National d'\'Etudes Spatiales
(CNES), which supported the PRONAOS project.
We are very grateful to the
PRONAOS technical teams at CNRS and CNES, and to the NASA-NSBF balloon-launching facilities group of Fort Sumner (New Mexico).

\end{document}